\documentclass[12pt]{article}

\usepackage{geometry}
\usepackage[utf8]{inputenc}
\usepackage[T1]{fontenc}
\usepackage[tt=false,type1=true]{libertine}
\usepackage[varqu]{zi4}
\usepackage[sort]{natbib}

\usepackage{hyperref}
\usepackage{url}
\usepackage{booktabs}
\usepackage{amsfonts}
\usepackage{nicefrac}
\usepackage{microtype}
\usepackage{xcolor}
\usepackage{multicol}
\usepackage{multirow}
\usepackage{subcaption}
\usepackage{mathrsfs}
\usepackage{geometry}
\usepackage[flushleft]{threeparttable}
\usepackage{amssymb,amsmath,latexsym,graphicx,amsthm}
\usepackage[libertine]{newtxmath}

\providecommand{\keywords}[1]
{
  \small
  \textbf{\textit{Keywords---}} #1
}

\title{Robust Auction Design in the Auto-bidding World}

\author{%
  Santiago Balseiro \\
  Columbia University and Google \\
  \texttt{srb2155@columbia.edu} \\
  \and
  Yuan Deng \\
  Google \\
  \texttt{dengyuan@google.com} \\
  \and
  Jieming Mao \\
  Google \\
  \texttt{maojm@google.com} \\
  \and
  Vahab Mirrokni \\
  Google \\
  \texttt{mirrokni@google.com} \\
  \and
  Song Zuo \\
  Google \\
  \texttt{szuo@google.com}
}
\newcommand{\pos}{\text{pos}}
\newcommand{\wel}{\text{Wel}}
\newcommand{\rev}{\text{Rev}}
\newcommand{\vcg}{\text{VCG}}
\newcommand{\gsp}{\text{GSP}}
\newcommand{\fpa}{\text{FPA}}

\newcommand{\opt}{\text{OPT}}

\newcommand{\xhdr}[1]{\noindent{\bf #1}}

\newcommand{\A}{\mathcal{A}}

\newtheorem{theorem}{Theorem}[section]
\newtheorem{lemma}[theorem]{Lemma}
\newtheorem*{theorem*}{Theorem}
\newtheorem*{lemma*}{Lemma}

\newtheorem{corollary}[theorem]{Corollary}

\newcommand\numberthis{\addtocounter{equation}{1}\tag{\theequation}}

\begin{document}

\maketitle

\begin{abstract}
In classic auction theory, reserve prices are known to be effective for improving revenue for the auctioneer against quasi-linear utility maximizing bidders. The introduction of reserve prices, however, usually do not help improve total welfare of the auctioneer and the bidders. In this paper, we focus on value maximizing bidders with return on spend constraints---a paradigm that has drawn considerable attention recently as more advertisers adopt auto-bidding algorithms in advertising platforms---and show that the introduction of reserve prices has a novel impact on the market. Namely, by choosing reserve prices appropriately the auctioneer can improve not only the total revenue but also the total welfare. Our results also demonstrate that reserve prices are robust to bidder types, i.e., reserve prices work well for different bidder types, such as value maximizers and utility maximizers, without using bidder type information.
We generalize these results for a variety of auction mechanisms such as VCG, GSP, and first-price auctions. Moreover, we show how to combine these results with additive boosts to improve the welfare of the outcomes of the auction further. Finally, we complement our theoretical observations with an empirical study confirming the effectiveness of these ideas using data from online advertising auctions.
\end{abstract}


\section{Introduction}

As auto-bidding---the practice of using optimization algorithms to procure advertising slots---is becoming the prevalent option in online advertising, a growing body of work is revisiting auction theory from the lens of the auto-bidding world~\citep{aggarwal2019autobidding,deng2021towards,balseiro2021landscape,babaioff2020non,golrezaei2021auctions}. A benefit of auto-bidding is that it simplifies advertisers' bidding process by asking advertisers for their high-level goals and then bidding on behalf of the advertisers.\footnote{See \href{https://support.google.com/google-ads/answer/6268637}{https://support.google.com/google-ads/answer/6268637} and \href{https://www.facebook.com/business/m/one-sheeters/value-optimization-with-roas-bidding}{https://www.facebook.com/business/m/one-sheeters/value-optimization-with-roas-bidding}; and \citet{aggarwal2019autobidding,balseiro2021landscape,deng2021towards} for more background introductions.} The main difference with classic auction theory stems from the model adopted for agent behavior. Unlike the classic {\em utility maximization} model, where each agent aims to maximize its own quasi-linear utility given by the difference between value and payment, the behavior of an auto-bidding agent is determined by its underlying optimization algorithm. In particular, the prevalent adopted model for the behavior of auto-bidding agents is that of {\em value maximization}, where each agent aims to maximize total values subject to an {\em ROS (return on spend) constraint} on the average spend per opportunity. For example, two common auto-bidding strategies are target CPA (cost per acquisition) and target ROAS (return on ad spend) auto-bidding, in which an algorithm optimizes the total {\em value} (e.g., the number of conversions) subject to an ROS constraint specified by the advertiser (e.g., the average spend for each conversion should not exceed a pre-specified target).

One surprising observation when agents are value maximizers is that the Vickrey-Clarke-Groves (VCG) auction, which is truthful and efficient for utility maximizers, is no longer truthful nor efficient. Value maximizers have incentives to strategize their bids and, moreover, there exist auction instances where the social welfare at equilibria under the VCG auction is only $1/2$ of the optimal social welfare~\citep{aggarwal2019autobidding,deng2021towards}. This is because the ratio between bids and values can be very different for each value-maximizing agent, and an agent with a low value may end up outbidding other bidders with high values in some auctions, creating allocation inefficiency. While there have been recent work addressing this problem, most previous studies have several shortcomings, e.g., they reduce such efficiency loss by introducing boosts that explicitly or implicitly rely on the accurate knowledge of the advertiser values~\citep{deng2021towards}, or they aim to optimize for either value-maximizing buyers or utility maximizing buyers (and not a mixture of auto-bidders)~\citep{balseiro2021landscape}, or they aim to optimize only welfare and not revenue~\citep{deng2021towards}.

In this paper, we aim to address the above shortcomings and propose simple auctions taking inaccurate value signals as additional inputs and we prove that the auctions have robust approximation guarantees in terms of both social welfare and revenue. Such inaccurate value signals could be the outcomes of some machine learning models that have bounded multiplicative errors on advertiser values.\footnote{The learning task of the signals can be a complex problem given the potential interaction with advertiser incentives. In this paper, we are agnostic about how the signals are learned and hence the learning problem is out of the scope. Nevertheless, the line of work on incentive-aware learning \citep{epasto2018incentive,golrezaei2021dynamic} could be relevant to avoid or mitigate inappropriate incentives.} Our key theoretical result is a generic lemma that transfers the accuracy of the value signals into approximation guarantees on both welfare and revenue for various auctions. Moreover, our main theorems not only work in the pure auto-bidding world, but also in mixed environments in which both value maximizers and utility maximizers coexist. Furthermore, the approximation guarantees apply to a very general set of market outcomes, i.e., they hold as long as no bidder is using a bidding profile that is always dominated under any competing bids. Such a broad solution concept includes classical notions such as complete-information Nash equilibrium as refinements and, as we discuss in our empirical results, additionally provides guarantees that hold along the convergence path when agents individually optimize their strategies using simple update rules.

Our auctions are based on VCG or generalized second-price auctions (GSP) and use the inaccurate signals as reserve prices and/or as (additive) boosts. Reserve prices are minimum prices the agents should clear to win the auction. Boosts are additive transformations on top of bids, i.e., agents are ranked based on the sum of their bids and boosts. Moreover, boosts are subtracted from standard payments to maintain incentive compatibility for utility maximizers. Intuitively, adding properly chosen boosts can push the auction to a more efficient outcome~\citep{deng2021towards}. On the other hand, 
in contrast to the classic utility maximization setup in which reserve prices usually do not help improve social welfare, modest reserve prices in fact may improve social welfare when there are value-maximizing agents in the market. The key difference is that in truthful auctions like the VCG auction, utility-maximizers never react to reserve prices, but value-maximizers do react to reserve prices to satisfy their ROS constraints. 
It turns out that, with properly chosen reserve prices, one can eliminate {\em bad} bidding strategies by reducing the set of bidding strategies that could be best responses to other bidders' strategies. As a result, the remaining {\em good} bidding strategies lead to an outcome with better social welfare guarantees.

\subsection{Our results}
We propose a set of simple auctions and prove approximation guarantees both in terms of social welfare and revenue. In particular, these approximation guarantees for the proposed auctions are all tight except for the social welfare approximation for GSP with reserve (Appendix~\ref{sec:tight_app}). The conclusions are {\em robust} against: i) signal inaccuracy, ii) agent behavior models (value/utility maximization or intermediate), iii) system status beyond equilibrium. The following Table~\ref{tab:results} summarizes most of our approximation results, where $\gamma \in [0, 1]$ describes the approximation accuracy of the signals and $\lambda \in [0, 1]$ indicates a hybrid behavior model between pure value maximization and pure utility maximization (e.g., $\lambda = 0$ for pure value maximization and $\lambda = 1$ for pure utility maximization). All formal definitions will be given later in Section~\ref{sec:prelim}.

\begin{table}
  \begin{threeparttable}
  \caption{Social welfare and revenue approximation guarantees for different auctions with $\gamma$-approximate value signals ($\gamma \in [0,1]$). For boosts we need: $\mathsf{signal \in [\gamma \cdot \nu \cdot \mathsf{value}, \nu \cdot \mathsf{value})}$ with $\nu \geq 1$; for reserves we need: $\mathsf{signal \in [\gamma \cdot \mathsf{value}, \mathsf{value})}$.}
  \label{tab:results}
  \centering
  \begin{tabular}{c  c  c  c  c}
    \toprule
    Auction & Behavior model\footnotemark[1] & Social welfare\footnotemark[2] & Revenue & Theorem  \\
    \midrule 
    VCG with reserve & $\lambda \in [0,1]$ & $1/(2-\gamma)$ & $\gamma$ & Corollary~\ref{cor:vcg_reserve}  \\
    VCG with boost & $\lambda \in [0,1]$ & $1/(2-\gamma)$ & - & Corollary~\ref{cor:vcg_boost}  \\
    VCG with reserve and boost & $\lambda \in [0,1]$ & $(1+\gamma)/2$ & $\gamma$ & Corollary~\ref{cor:vcg_reserve_boost}  \\
    GSP with reserve & $\lambda \in [0,1]$ & $\gamma$ & $\gamma$ & Corollary~\ref{cor:gsp}  \\
    GSP with reserve and boost & $\lambda = 0$ & $(1+\gamma)/2$ & $\gamma$ & Corollary~\ref{cor:gsp_u}  \\
    \bottomrule
  \end{tabular}
  \begin{tablenotes}
    \small
    \item[1] $\lambda = 0$: value maximization; $\lambda = 1$: utility maximization; $\lambda \in (0,1)$: intermediate models (see Program~\eqref{eq:behavior}).
    \item[2] For $0 < \gamma < 1$, $(1+\gamma)/2 > 1/(2-\gamma) > \gamma$.
  \end{tablenotes}
  \end{threeparttable}
\end{table}

Observe that in a world with value maximizers only (i.e., $\lambda = 0$), if all the value maximizers can hit their targets (see Program \eqref{eq:behavior}), the revenue equals to the social welfare. In such an environment, our social welfare guarantees can directly imply the same revenue guarantees under the condition that value maximizers all hit their targets.

We provide a general framework for proving approximation results in the presence of value maximizers through a novel technical lemma (Lemma~\ref{lem:meta_boost}). In fact, our approximation results are mostly derived from this lemma. Finally, we conduct empirical analyses with semi-synthetic data and validate our theoretical findings for the performance of our mechanisms in VCG auctions.

\subsection{Related Work}

As a central topic in economic study, since the seminal work of Vickrey–Clarke–Groves (VCG) auctions~\citep{vickrey1961counterspeculation,clarke1971multipart,groves1973incentives} and Myerson's auction~\citep{myerson1981optimal}, auction design has been successfully deployed in many different fields. Examples includes combinatorial auctions for reallocating radio frequencies~\citep{cramton2006combinatorial} and generalized second-price auctions as well as dynamic auctions for online advertising~\citep{aggarwal2006truthful,edelman2007internet,varian2007position,mirrokni2020non}. 

In contrast to works assuming utility-maximizing agents, a growing body of work has recently focused on auto-bidders such as target CPA bidders and target ROAS bidders. \citet{aggarwal2019autobidding} find the optimal bidding strategies for a general class of auto-bidding objectives and prove the existence of pure-strategy equilibrium. \citet{deng2021towards} show how boosts can be used to improve the efficiency guarantees when target CPA and target ROAS auto-bidders coexist. \citet{balseiro2021landscape} characterize the revenue-optimal single-stage auctions with either value-maximizers or utility-maximizers with ROS constraints under various information structure. \citet{golrezaei2021auctions} study the auction design for utility-maximizers with ROI constraints, where the behavior model is equivalent to the special case of ours with $\lambda = 1 / (1 + \mathsf{minimal~ROI})$. Besides the model of constrained optimization, there is another generalization of the utility models with ROI constraints adopted by \citet{goel2014clinching,goel2019pareto,babaioff2020non}. In particular, the negative payment term in the utility function is replaced by a general cost, which is often a convex function of the payment.

Independent of the above, the term of value-maximizer has been used in some other works but with quite different mathematical models \cite{fadaei2016truthfulness,wilkens2016mechanism,wilkens2017gsp}. The main distinction is that in these models, the target ROS constraints are imposed on a single auction rather than across a set of auctions, hence the bidders become insensitive to the marginal tradeoff between values and payments, leading to different behavior patterns.

Prior to these, the simplest auto-bidding model is budget optimization with utility-maximizers~\citep{borgs2007dynamics}. \citet{pai2014optimal} characterize the revenue-optimal auction of utility-maximizing bidders with budget constraints. For applications in online advertising, \citet{balseiro2019learning} develop budget management strategies that are no-regret in a long run, which is extended in more complex settings \citep{pmlr-v119-balseiro20a,pmlr-v139-balseiro21a,avadhanula2021stochastic,celli2021parity}. \citet{balseiro2017budget,balseiro2020budget} provide a thorough study to compare different commonly used budget management strategies in practice. \citet{conitzer2017multiplicative,conitzer2018pacing} study pacing algorithms for budget constraints in both first price auctions and second price auctions.

The techniques of reserve prices and additive boosts have been widely studied in the literature~\citep{amin2013learning,paes2016field,sandholm2015automated,lavi2003towards,deng2021revenue,golrezaei2021boosted}. Among these works, \citet{deng2021towards} improve the welfare approximation ratio to $(c+1) / (c+2)$ using boosted auctions with accurate signals about bidders' values in the environment with value maximizers only. Our results are more robust and more general from several aspects: We allow inaccurate signals that approximate bidders' values; Our results hold for a mixture of behavior models in auctions with reserves and/or boosts; We provide guarantees on revenue performance in addition to welfare performance. 

Some of our results rely on the assumption that the auto-bidders adopt uniform bidding strategies when the underlying auction is not incentive-compatible for utility-maximizing bidders. In practice, it is usually hard for bidders to adopt and optimize non-uniform bidding strategies, and moreover, uniform bidding has been shown to perform well against optimal non-uniform bidding strategies in ad auctions~\citep{feldman2007budget,feldman2008algorithmic,bateni2014multiplicative,balseiro2019learning,deng2020data}.
\section{Preliminaries}\label{sec:prelim}

\xhdr{Position Auctions.}
We consider a setting with $n$ bidders bidding simultaneously in $m$ position auctions~\citep{lahaie2007sponsored,varian2007position}. 
In each position auction $j$, we have $s_j$ slots which can be allocated to $s_j$ different bidders. For each bidder $i \in [n]$, we use $v_{i,j}$ to denote the base value of auction $j$ for bidder $i$, and the value of the $k$-th slot in auction $j$ is $v_{i,j} \cdot \pos_{j,k}$. Here $\pos_{j,k} \in \mathbb{R}^+$ is the position normalizer of the $k$-th slot in auction $j$ which does not depend on $i$. Without loss of generality, we assume $\pos_{j,k}$ is decreasing in $k$. 
For notation convenience, we set $\pos_{j,s_j+1} = 0$ for the non-existing slot. We use $I = (n,m,\{s\}_j, \{v\}_{i,j}, \{\pos\}_{j,k})$ to denote a problem instance and ${}_{-i}$ to denote the bidders other than bidder $i$. The optimal welfare of the problem instance is defined as 
\[
\wel(\opt) = \sum_{j=1}^m \sum_{k =1}^{s_j} \pos_{j, k}\cdot k\text{-th highest value of } (v_{1,j}, ..., v_{n,j}).
\]

We use $b, p, x$ to denote bidders' bids, payments and allocations. More particularly, $b_{i, j}$ is bidder $i$'s bid in auction $j$. $x_{i,j,k}$ is 1 if bidder $i$ gets the $k$-th slot of auction $j$ and 0 otherwise. $p_{i,j}$ is the price paid by the bidder $i$ in auction $j$. The welfare and revenue of an allocation $x$ and payments $p$ are defined as
\[
\wel(x,p) = \sum_{i=1}^n \sum_{j=1}^m \sum_{k =1}^{s_j} x_{i,j,k} \cdot v_{i,j}\cdot \pos_{j,k},  
\qquad \text{and} \qquad
\rev(x,p) = \sum_{i=1}^n \sum_{j=1}^m p_{i,j}.
\]
For notation convenience, we will use $\wel_i(x,p) =  \sum_{j=1}^m \sum_{k =1}^{s_j} x_{i,j,k} \cdot v_{i,j}\cdot \pos_{j,k} $ and $\rev_i(x,p)=\sum_{j=1}^m p_{i,j}$ to represent each bidder's contribution to welfare and revenue.

In this paper, we mainly focus on three auction formats used in position auctions: the Vickrey–Clarke–Groves auction (VCG), the generalized second-price auction (GSP), and the first-price auction (FPA). Their allocation rules are the same: rank bidders by their bids (tie-breaking by bidder indices) and allocate the $k$-th slot to the bidder with the $k$-th highest bid. In the VCG auction each agent pays the externality it imposes on the other the agents, in the GSP auction each agent who is allocated pays the bid of the next highest bidder, and in the FPA each agent who is allocated pays their bid. For notation convenience, we denote by $\hat{b}_{k,j}$ the $k$-th highest bid for $k \in [s_j]$ in auction $j$. Assuming bidder $i$ wins slot $k$ in auction $j$, its payment $p_{i,j}$ in the three auction formats are: (1) VCG: $p_{i,j} = \sum_{\kappa = k+1}^{s_j} \hat{b}_{\kappa,j}\cdot (\pos_{j, \kappa-1} - \pos_{j, \kappa})$, (2) GSP: $p_{i,j} = \hat{b}_{k+1,j} \cdot \pos_{j,k}$, and (3) FPA: $p_{i,j} = b_{i,j} \cdot \pos_{j,k}$. It is not hard to see that, with the same bids, the payments from these three auctions is ranked in an increasing order as VCG, GSP, FPA. 

\xhdr{Reserve Prices and Boosts.}
We further consider three auction formats with reserve prices and boosts. We denote by $r_{i,j}$ the reserve price and by $z_{i,j}$ the boost for bidder $i$ in auction $j$. 
When we have boosts, bidders are ranked by their score $b_{i,j}+z_{i,j}$ and we use $\hat{b}_{k,j}$ to denote the $k$-th highest score for $k \in [s_j]$ in auction $j$. For ease of presentation, we consider lazy reserves such that the slot is not allocated to a bidder if her bid without boosts does not clear her reserve, i.e., $b_{i,j} < r_{i,j}$; our results continue to hold for eager reserves~\citep{paes2016field}.
With reserve prices and boosts, the prices of VCG, GSP and FPA when bidder $i$ gets slot $k$ in auction $j$ become:
\begin{itemize}
    \item VCG: $p_{i,j} = \sum_{\kappa = k+1}^{s_j} \max(\hat{b}_{\kappa,j}-z_{i,j}, r_{i,j})\cdot (\pos_{j, \kappa-1} - \pos_{j, \kappa})$;
    \item GSP: $p_{i,j} = \max(\hat{b}_{k+1,j}-z_{i,j}, r_{i,j}) \cdot \pos_{j,k}$;
    \item FPA: $p_{i,j} = (\hat{b}_{i,j}-z_{i,j}) \cdot \pos_{j,k}$.
\end{itemize}

It is worth highlighting that boosts and reserves are fairly different components in auctions. Reserves filter out the candidates whose bids are lower than their reserves and then the remaining candidates are ranked according to their original bids. In contrast, boosts are added to the candidates’ ranking scores, and therefore, the candidates are ranked according to their original bids plus their boosts.

\xhdr{Inaccurate signals.}
Our auctions take inaccurate value signals as input and use them as reserve prices and/or boosts. In particular, for $\gamma \in [0, 1]$: 
\begin{itemize}
    \item When the signals are used as boosts, we allow multiplicative errors in both directions and we say boosts are $\gamma$-approx, if for any $i \in [n],j \in [m]$, $z_{i,j} \in [\mu\cdot v_{i,j},\nu\cdot v_{i,j})$, and $\mu = \gamma \cdot \nu$;
    \item When the signals are used as reserve prices, we allow multiplicative underestimation errors and say reserve prices are $\gamma$-approx, if for any $i \in [n],j \in [m]$, $r_{i,j} \in [\gamma\cdot v_{i,j}, v_{i,j})$.
\end{itemize}
Note that we require a one-direction error for reserve signals. One can always convert signals with bounded errors in both directions into signals with only underestimation errors by scaling them down once multiplicative overestimation errors have a finite upper bound.

\xhdr{Bidders.}
We focus on two types of bidders: utility maximizers and value maximizers. A utility maximizer bidder $i$ maximizes
$\wel_i(x,p) - \rev_i(x,p)$.
In contrast, a value maximizer maximizes
$\wel_i(x,p)$ subject to a return on spend constraint $\wel_i(x,p) \geq \rev_i(x,p)$. Here, it is without loss of generality to assume the target ratio between return $\wel_i(x,p)$ and spend $\rev_i(x,p)$ is $1$.

These two types of bidders can be summarized by bidders optimizing the following program
\begin{align}\label{eq:behavior}
  \max \qquad & \wel_i(x,p) - \lambda_i \cdot \rev_i(x,p)  \\
  \text{s.t.} \qquad & \wel_i(x,p) \geq \rev_i(x,p). \nonumber
\end{align}
Here $\lambda_i = 0$ corresponds to a value maximizer and $\lambda_i = 1$ corresponds to a utility maximizer. Our results apply for bidders with $\lambda_i \in [0,1]$. We allow $\lambda_i$ to be different for each bidder and they are unknown to the auctioneer. Here, utility maximizers can be modeled with $\lambda = 1$ because all the auctions we consider are individual rational. As a result, the optimal solution of the objective in \eqref{eq:behavior} is always non-negative, and therefore, the constraint would be irrelevant for utility maximizers.

\xhdr{Solution Concept.} We consider a solution concept called \emph{undominated bids} which includes the support of Nash equilibrium bids as a subset. Because our results apply to the larger set of undominated bids, they readily hold for refinements such as Nash equilibria, in which our results give price of anarchy bounds~\citep{papadimitriou2001algorithms}.
Following the standard definition of (weak) dominance, for a problem instance $I$ and an auction format, we say a bid vector $b_i = (b_{i,1}, ...,b_{i,m})$ is (weakly) dominated by another bid vector $b'_i = (b'_{i,1}, ... ,b'_{i,m})$, if the following two requirements are satisfied:

Let $x,p$ be the allocation and prices induced by $b_i ,b'_{-i}$, and $x',p'$ be the allocation and prices induced by $b'_i ,b'_{-i}$. Then
for any other bidders' bids $b'_{-i}$, $b'_i$ is at least as good as $b_i$, i.e.,
\begin{itemize}
    \itemsep0em
    \item both violate constraints: $\wel_i(x,p) < \rev_i(x,p)$ and $\wel_i(x',p') < \rev_i(x',p')$
    \item $b_i$ violates constraints while $b'_i$ does not:
    $\wel_i(x,p) < \rev_i(x,p)$ and $\wel_i(x',p') \geq \rev_i(x',p')$
    \item or neither violates constraints, and $b'_i$ yields no worse objective: $\wel_i(x,p) \geq \rev_i(x,p)$ and $\wel_i(x',p') \geq \rev_i(x',p')$ and $\wel_i(x',p') -\lambda_i\cdot \rev_i(x',p') \geq \wel_i(x,p) -\lambda_i\cdot \rev_i(x,p)$
\end{itemize}
There exits $b'_{-i}$ such that $b'_i$ is strictly better than $b_i$, i.e.,
\begin{itemize}
    \setlength\itemsep{0em}
    \item $b_i$ violates constraints while $b'_i$ does not: $\wel_i(x,p) < \rev_i(x,p)$ and $\wel_i(x',p') \geq \rev_i(x',p')$
    \item or neither violates constraints, and $b'_i$ yields strictly better objective: $\wel_i(x,p) \geq \rev_i(x,p)$ and $\wel_i(x',p') \geq \rev_i(x',p')$ and $\wel_i(x',p') -\lambda_i\cdot \rev_i(x',p') > \wel_i(x,p) -\lambda_i\cdot \rev_i(x,p)$
\end{itemize}

We omit the word ``weak'' for convenience for the rest of the paper. We say $b_i$ is undominated if there is no $b'_i$ dominates $b_i$. Denote the set of all bidders' bids $b = (b_1,..., b_n)$ by $\Theta$ in which each $b_i$ is undominated for $i \in [n]$ and each bidder is paying at most its welfare.

As uniform bidding is widely adopted in automated bidding strategies, we also consider a solution concept related to it. We say $b_i$ is a uniform bidding for bidder $i$, if $b_{i,j} = v_{i,j} \cdot \delta_i$ for all $j \in [m]$. We use $\Theta_u$ to denote the set of all bidders' bids $b = (b_1,...,b_n)$ such that each $b_i$ is a uniform bidding and is not dominated by any other uniform bidding and each bidder is paying at most its welfare.

\section{Main Technical Lemma}
\label{sec:main_lemma}

We first show a lemma which will be used as the major technical building block for our results. This lemma provides a general framework for proving approximation results on revenue and welfare. Informally, this lemma says that once we can guarantee lower bounds on bidders' bids and apply reserve prices and/or boosts, these immediately lead to lower bound guarantees on revenue and welfare in terms of the optimal welfare. Notice that this lemma continues to hold when we apply reserve prices only (i.e., $\mu=\nu=0$) or boosts only (i.e., $\beta = 0$). For a better understanding of the lemma statement, conditions number 1, 2 and 5 hold by the definition of the model and conditions number 3 and 4 will be proved hold when using this lemma.
\begin{lemma}
\label{lem:meta_boost}
Consider running $m$ position auctions with allocation and pricing rule $\A$. Assuming $\A$ together with bidders' bids $b$ satisfy the following conditions for parameters $\alpha,\beta,\mu,\nu \geq 0$:
\begin{itemize}
    \item For each bidder $i$ and each auction $j$, the reserve satisfies $r_{i,j} \geq \beta \cdot v_{i,j}$ and the boost satisfies $\mu\cdot v_{i,j}  \leq z_{i,j} < \nu\cdot v_{i,j}$.
    \item In each position auction, bidders are ranked by their scores $b_{i,j}+z_{i,j}$ and the $k$-th highest score wins the $k$-th slot if $b_{i,j} \geq r_{i,j}$. 
    \item If bidder $i$'s value $v_{i,j}$ ranks in top-$s_j$ in auction $j$, bidder $i$ bids at least $\alpha$ times its base value, i.e. $b_{i,j} \geq \alpha \cdot v_{i,j}$. 
    \item If bidder $i$ wins slot $k$ in auction $j$, bidder $i$ pays at least the VCG price, i.e. $p_{i,j} \geq  \sum_{\kappa = k+1}^{s_j} (\pos_{j, \kappa-1} - \pos_{j,\kappa}) \cdot \max(\hat{b}_{\kappa,j} -z_{i,j}, r_{i,j})$.
    \item For each bidder, her total payment is at most her total value. 
\end{itemize}

We have $\rev(\A(b)) \geq \min\left(\frac{(\alpha +\mu)\beta}{\beta+\nu}, \beta\right) \cdot\wel(\opt)$ and $\wel(\A(b)) \geq\frac{\alpha+\mu}{1 + \max(\nu, \alpha+ \mu -\beta)} \cdot\wel(\opt)$.
\end{lemma}

We give a high-level proof sketch of this lemma. The detailed proof can be found in Appendix \ref{sec:main_lemma_app}.

\xhdr{Proof Sketch.} Informally, the first step of the proof is to fractionally partition the auction slots into two parts so that (1) In part A, $\A(b)$ and $\opt$ agree in allocation; (2) In part B, $\A(b)$ and $\opt$ disagree in allocation. We then lower bound the welfare and the revenue of $\A(b)$ in these two parts separately with different arguments.

In part A, $\A(b)$ and $\opt$ have the same allocation. The welfare of $\A(b)$ is by definition the same as the welfare of $\opt$. Moreover, the revenue of $\A(b)$ can be lower bounded by $\beta$ times the welfare of $\opt$, via conditions on reserve prices (the first bullet point of the lemma statement).

In part B, $\A(b)$ and $\opt$ have different allocations as allocated bidders of $\opt$ are not allocated in $\A(b)$. Since we have lower bounds on bids (the third bullet point of the lemma statement), these bidders' bids would give a lower bound on the revenue of $\A(b)$ via VCG pricing. The additional boost $z$ (in the first bullet point of the lemma statement) would escalate this effect and give a lower bound on a linear combination of the welfare and the revenue of $\A(b)$.

By putting these lower bounds together with the condition that $\wel(\A(b)) \geq \rev(\A(b))$ (in the fifth bullet point of the lemma statement), we can obtain lower bounds on $\wel(\A(b))$ and $\rev(\A(b))$.

\section{Applications in Auctions}
\subsection{VCG Auctions}
\label{sec:vcg}
In this section, we consider VCG with reserves and boosts. We first show the following lemma about the set of undominated bids $\Theta$. Its proof can be found in Appendix \ref{sec:vcg_app}.

\begin{lemma}
\label{lem:vcg}
For any problem instance $I$, let $\Theta$ be the set of undominated bids for VCG with reserve $r$ and boost $z$. Assume reserve $r$ satisfies that $r_{i,j}< v_{i,j} ~\forall i\in[n], j\in[m]$, and boost $z$ satisfies that $z_{i,j} \in [\mu\cdot v_{i,j}, \nu\cdot v_{i,j})$ for some $\nu -\mu \leq 1$ for all $i\in[n], j\in[m]$. For any $b \in \Theta$, we have $b_{i,j} \geq v_{i,j}$, if bidder $i$'s value $v_{i,j}$ ranks in top-$s_j$ in auction $j$.
\end{lemma}

With Lemma \ref{lem:vcg}, we are ready to state our results on VCG with reserves and boosts. The following three corollaries are the results we get when we apply (1) only reserve prices (2) only boosts (3) reserve prices and boosts together. 

Combining Lemma \ref{lem:vcg} with Lemma \ref{lem:meta_boost} using $\alpha = 1$, $\beta = \gamma$, and $\mu=\nu=0$: 
\begin{corollary}
\label{cor:vcg_reserve}
On any problem instance $I$, VCG with $\gamma$-approx reserve, denoted by $\vcg^{\gamma}_r$, satisfies 
\[
\wel(\vcg^{\gamma}_r(b)) \geq \frac{1}{2-\gamma} \cdot \wel(\opt)\text{~~~~and~~~~}\rev(\vcg^{\gamma}_r(b)) \geq \gamma \cdot \wel(\opt),
\]
for bids $b$ from the undominated bids set $\Theta$.
\end{corollary}
Combining Lemma \ref{lem:vcg} with Lemma \ref{lem:meta_boost} using $\alpha = 1$, $\beta = 0$, $\mu = \gamma/(1-\gamma)$, and $\nu = 1/(1-\gamma)$: 
\begin{corollary}
\label{cor:vcg_boost}
On any problem instance $I$, VCG with $\gamma$-approx boosts, denoted by $\vcg^{\gamma}_b$, satisfies 
\[
\wel(\vcg^{\gamma}_b(b)) \geq \frac{1}{2-\gamma} \cdot \wel(\opt)
\]
for bids $b$ from the undominated bids set $\Theta$.
\end{corollary}
Combining Lemma \ref{lem:vcg} with Lemma \ref{lem:meta_boost} using $\alpha = 1$, $\beta = \gamma$, $\mu = \gamma$, and $\nu = 1$: 
\begin{corollary}
\label{cor:vcg_reserve_boost}
On any problem instance $I$, VCG with $\gamma$-approx reserve and $\gamma$-approx boost, denoted by $\vcg^{\gamma}_{r,b}$, satisfies 
\[
\wel(\vcg^{\gamma}_{r,b}(b)) \geq \frac{\gamma+1}{2} \cdot \wel(\opt)\text{~~~~and~~~~}\rev(\vcg^{\gamma}_{r,b}(b)) \geq \gamma \cdot \wel(\opt),
\]
for bids $b$ from the undominated bids set $\Theta$.
\end{corollary}
\subsection{Generalized Second-Price Auctions}
\label{sec:gsp}
For GSP, we are able to show a lemma similar to Lemma \ref{lem:vcg} assuming bidders are all value maximizing ($\lambda_i = 0~ \forall i\in[n]$) and they are uniform bidding. Its proof can be found in Appendix \ref{sec:gsp_app}.
\begin{lemma}
\label{lem:gsp_u}
For any problem instance $I$ with $\lambda_i = 0~ \forall i\in[n]$, let $\Theta_u$ be the set of undominated uniform bids for GSP with with reserve $r$ and boost $z$. Assume reserve $r$ satisfies that $r_{i,j}< v_{i,j} ~\forall i\in[n], j\in[m]$, and boost $z$ satisfies that for some $\nu -\mu \leq 1$, $z_{i,j} \in [\mu\cdot v_{i,j}, \nu\cdot v_{i,j})~\forall i\in[n], j\in[m]$. For any $b \in \Theta_u$, we have $b_{i,j} \geq v_{i,j}$, if bidder $i$'s value $v_{i,j}$ ranks in top-$s_j$ in auction $j$.
\end{lemma}

Combining Lemma \ref{lem:gsp_u} with Lemma \ref{lem:meta_boost}, we can obtain results similar to Corollary \ref{cor:vcg_reserve},\ref{cor:vcg_boost}, \ref{cor:vcg_reserve_boost}. We only state the strongest one with both boosts and reserve prices: 
\begin{corollary}
\label{cor:gsp_u}
On any problem instance $I$ with $\lambda_i = 0~ \forall i\in[n]$, GSP with $\gamma$-approx reserve and $\gamma$-approx boost, denoted by $\gsp^{\gamma}_{r,b}$, satisfies 
\[
\wel(\gsp^{\gamma}_{r,b}) \geq \frac{\gamma+1}{2} \cdot \wel(\opt)\text{~~~~and~~~~}\rev(\gsp^{\gamma}_{r,b}(b)) \geq \gamma \cdot \wel(\opt),
\]
for uniform bids $b$ from the undominated uniform set $\Theta_u$.
\end{corollary}

When there are no restrictions on bidders' bidding behavior, we can obtain the following weaker lemma for GSP with reserves. Its proof can be found in Appendix \ref{sec:gsp_app}.
\begin{lemma}
\label{lem:gsp}
For any problem instance $I$, let $\Theta$ be the set of undominated bids for GSP with $\gamma$-approx reserve $r$ and no boosts. For any $b \in \Theta$, we have $b_{i,j} \geq r_{i,j}\geq \gamma \cdot v_{i,j}$, $\forall i \in [n], j \in [m]$.
\end{lemma}

Combining Lemma \ref{lem:gsp} with Lemma \ref{lem:meta_boost} using $\alpha = \gamma$, $\beta = \gamma$, and $\mu = \nu = 0$:
\begin{corollary}
\label{cor:gsp}
On any problem instance $I$, GSP with $\gamma$-approx reserve, denoted by $\gsp^{\gamma}_r$, satisfies 
\[
\wel(\gsp^{\gamma}_r(b)) \geq \gamma \cdot \wel(\opt)\text{~~~~and~~~~}\rev(\gsp^{\gamma}_r(b)) \geq \gamma \cdot \wel(\opt),
\]
for bids $b$ from the undominated bids set $\Theta$.
\end{corollary}

\subsection{First-Price Auctions}
\label{sec:fpa}

For FPA, if we restrict that bidders are only value maximizers and they are uniform bidding, we know from prior work that optimal welfare and revenue can be achieved without reserves and boosts. Its proof can be found in Appendix~\ref{app:fpa}.
\begin{theorem*}[\cite{deng2021towards}]
On any problem instance $I$ with $\lambda_i = 0~ \forall i\in[n]$, FPA has 
\[
\wel(\fpa(b))=\rev(\fpa(b))=\wel(\opt),
\]
for uniform bids $b$ from the undominated uniform set $\Theta_u$.
\end{theorem*}

When there are no restrictions on bidders, we get similar results as GSP with the following lemma.

\begin{lemma}
\label{lem:fpa}
For any problem instance $I$, let $\Theta$ be the set of undominated bids for FPA with $\gamma$-approx reserve $r$ and no boosts. For any $b \in \Theta$, we have $b_{i,j} \geq r_{i,j} \geq \gamma \cdot v_{i,j}$, $\forall i \in [n], j \in [m]$.
\end{lemma}

Combining Lemma \ref{lem:fpa} with Lemma \ref{lem:meta_boost} using $\alpha = \gamma$, $\beta = \gamma$, and $\mu=\nu =0$:
\begin{corollary}
On any problem instance $I$, FPA with $\gamma$-approx reserve ($\fpa^{\gamma}_r$) has 
\[
\wel(\fpa^{\gamma}_r(b)) \geq \gamma \cdot \wel(\opt)\text{~~~~and~~~~}\rev(\fpa^{\gamma}_r(b)) \geq \gamma \cdot \wel(\opt),
\]
for bids $b$ from the undominated bids set $\Theta$.
\end{corollary}

\section{Experiments}

In this section, we derive semi-synthetic data from real auction data of a major search engine to validate our theoretical findings concerning VCG auctions. VCG auctions provide a clean environment for us to validate our findings as uniform bidding is a best response for each value maximizer~\citep{aggarwal2019autobidding} while computing best response in GSP auctions is a highly non-trivial task.

Observe that when the bidders are symmetric such that their valuation distributions are identically and independently distributed, both optimal efficiency and optimal revenue are achieved in any symmetric equilibrium, and thus, no efficiency or revenue improvement can be observed by applying our mechanisms. Therefore, we use real ad auction data for capturing variation across bidders. For the experimental purpose, we simulate VCG auctions with bids from value maximizers only, as utility maximizers do not respond to either boosts or reserve prices in VCG mechanisms. Instead of using real return on ad spend targets for value maximizers, we generate artificial targets to exclude any practical noises from the real system.
We emphasize that the main objective for our empirical study is
to validate our theoretical findings rather than investigating the
efficiency and/or revenue potentials on real systems actually implemented in
practice. Many practical aspects often need to be taken care of in the design of real systems, which would never be considered in theory.

\xhdr{Simulation Procedures.} To properly evaluate the efficiency and revenue of a new mechanism, after the generation of the dataset, we first pre-train the (uniform) bid multipliers $\delta$ for value maximizers in $25$ iterations without reserve prices and boosts to obtain an equilibrium as a starting point. We then simulate the response of value maximizers by gradient descent on their bid multipliers in log space until convergence~\citep{nesterov2013introductory,aggarwal2019autobidding}. Formally, let $\delta_{i, t}$ be the bid multiplier for value maximizer $i$ in iteration $t$. In addition, let $\wel_{i, t}$ be value maximizer $i$'s total received value in iteration $t$, and let $\rev_{i, t}$ be her total payment in iteration $t$. Then, in iteration $t+1$, the value maximizer $i$'s bid multiplier is updated by $\log \mathsf{\delta}_{i, t+1} = (1-\eta_t) \cdot \log \mathsf{\delta}_{i, t} + \eta_t \cdot \log \frac{\wel_{i, t}}{\rev_{i, t}}$,
where $\eta_t \in (0, 1)$ is a properly chosen learning rate for $t$-th iteration. Intuitively, when $\rev_{i, t} < \wel_{i, t}$, value maximizer $i$'s bid multiplier increases for the next iteration; otherwise, her bid multiplier decreases for the next iteration.
After obtaining a starting point, we simulate another $25$ iterations for auctions with reserve prices and/or boosts. In this way, we can observe both the initial impact of adding reserve prices and/or boosts and how the impact changes over time during value maximizers' response until convergence.

\xhdr{Reserve Prices and Boosts.} In addition to the baseline in which we continue to use auctions without reserve prices, we experiment with boosts and reserve prices using signals with different approximation factors $\gamma$. For each bidder $i$ in each auction $j$, we will set reserve prices or give boosts based on $\gamma$ in different treatments. Here, let $s_{i,j}^\gamma$ be a random variable independently sampled from a Gaussian distribution with mean $(1+\gamma) / 2$ and standard error $0.01$, truncated within $[\gamma, 1]$. In treatment $\text{reserve-}\gamma$, we set the reserve price as $s_{i,j}^\gamma \cdot v_{i,j}$. In treatment $\text{boost-}\gamma$, we add an additive boost $\frac{1}{1 - \gamma} \cdot s_{i,j}^\gamma \cdot v_{i,j}$ as suggested. Finally, in treatment $\text{boost-reserve-}\gamma$, we set the reserve price as $s_{i,j}^\gamma \cdot v_{i,j}$ and additionally give an additive boost $\frac{1}{1 - \gamma} \cdot s_{i,j}^\gamma \cdot v_{i,j}$. All metrics we report are relative to the gap between the baseline (i.e., without treatment) and the optimal solution. More precisely, let $\kappa_{\text{init}}$ be the initial welfare (revenue) without reserve prices or boosts, $\kappa_{\text{e}}$ be the welfare (revenue) under treatment $e$, and $\kappa(\mathsf{OPT})$ be the optimal welfare (revenue). Then the reported percentage is computed by $(\kappa_{\text{e}} - \kappa_{\text{init}}) / (\kappa(\mathsf{OPT}) - \kappa_{\text{init}})$.

\subsection{Experimental Results}

\begin{figure}[h]
    \centering
    \begin{subfigure}[b]{0.47\textwidth}
        \centering
        \includegraphics[width=\textwidth]{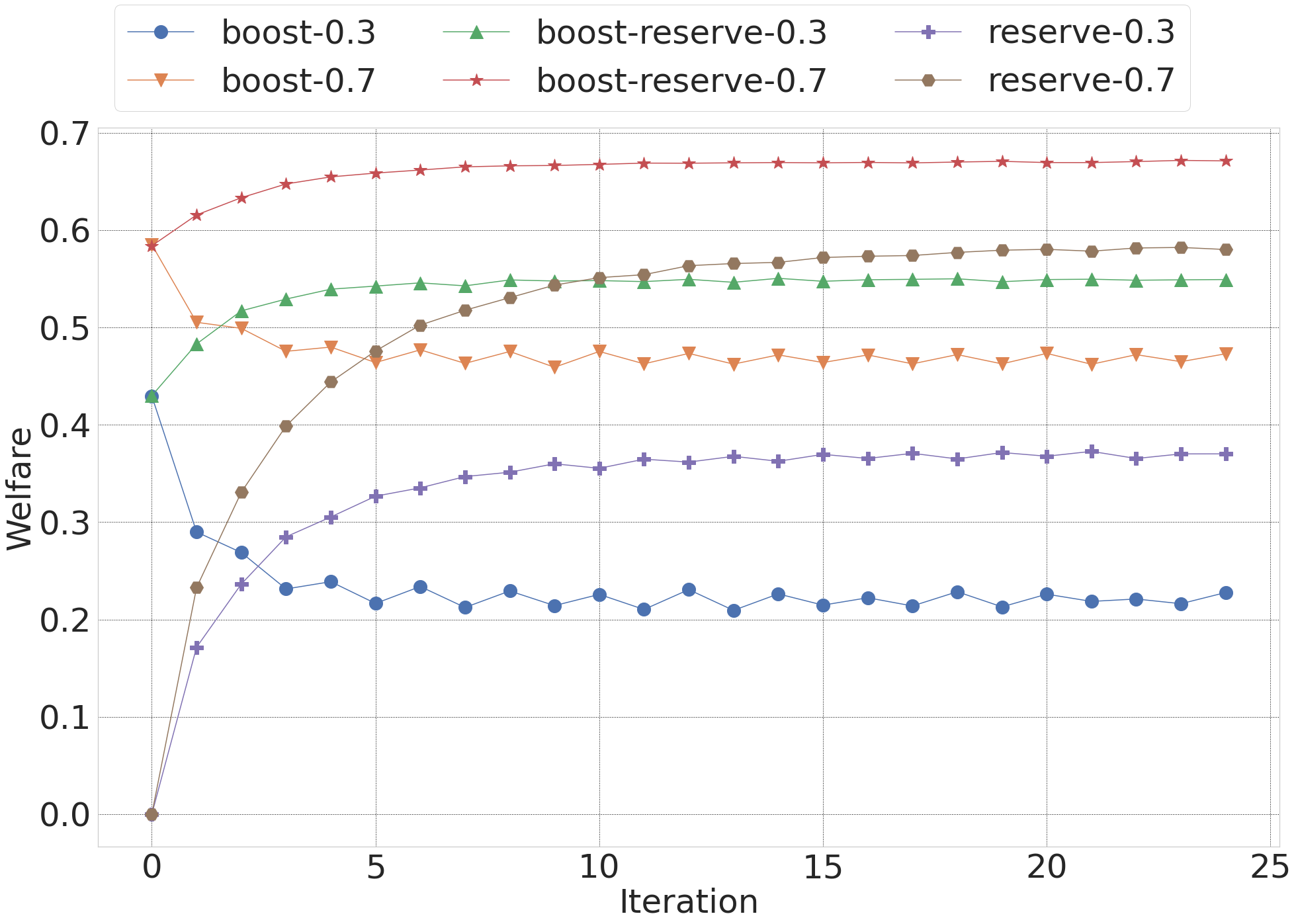}
        \caption{Trend of Welfare}
        \label{fig:welfare_trend}
    \end{subfigure}
    \qquad
    \begin{subfigure}[b]{0.47\textwidth}
        \centering
        \includegraphics[width=\textwidth]{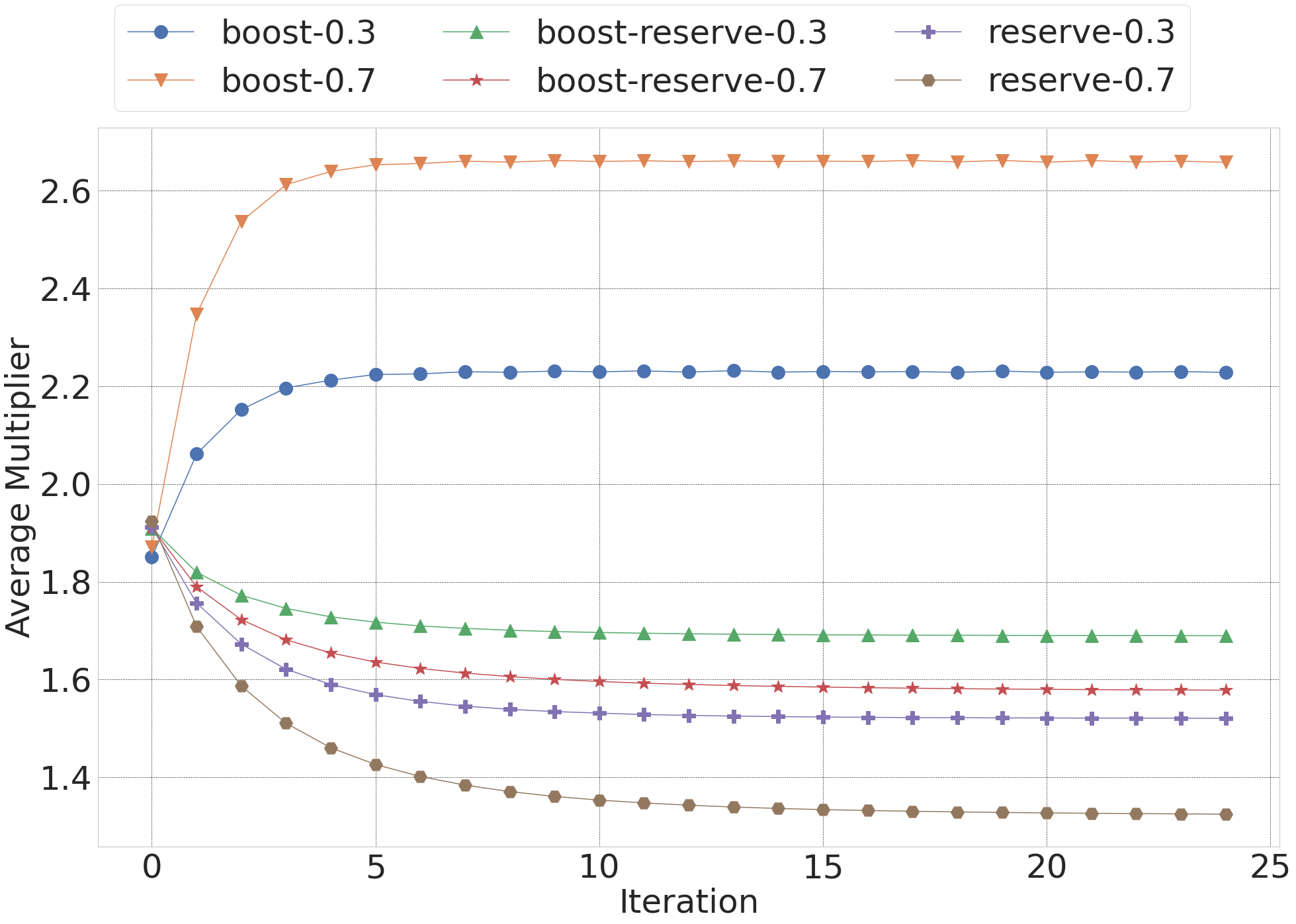}
        \caption{Trend of Average Bid Multiplier}
        \label{fig:multiplier_trend}
    \end{subfigure}
    \caption{Convergence during response per iteration.}
\end{figure}

Figure~\ref{fig:welfare_trend} reports the trend of welfare performance in VCG auctions under different treatments with one run of the experiment. Observe that both variants of reserve prices have neutral initial impact on welfare before value maximizers start to respond. After the value maximizers start to respond, they adjust their bid multipliers towards a better allocation, resulting in a positive impact on welfare. As the reserve prices become more precise (i.e., when $\gamma$ is 
closer to $1$), the welfare impact is larger, which confirms our theoretical results of Corollary~\ref{cor:vcg_reserve}. In contrast, both variants of boosts have positive initial impact on welfare. However, as the value maximizers start to respond, the welfare impact starts to decrease but the final impact after convergence is still positive, confirming our theoretical results of Corollary~\ref{cor:vcg_boost}. Interestingly, when treatment $\text{reserve-}\gamma$ and treatment $\text{boost-}\gamma$ share the same $\gamma$, their final impact on revenue after response are close to each other, as predicted by the same welfare bounds from Corollary~\ref{cor:vcg_reserve} and~\ref{cor:vcg_boost}. Moreover, a combination of boosts and reserve prices outperforms the treatments with either reserve prices only or boosts only, which validates Corollary~\ref{cor:vcg_reserve_boost} that enjoys a better bound than Corollary~\ref{cor:vcg_reserve} and~\ref{cor:vcg_boost}.

Figure~\ref{fig:multiplier_trend} demonstrates the trend of average (uniform) bid multipliers for value maximizers under different treatments in VCG auctions with one run of the experiment. We observe that the average multiplier decreases during response for both treatments with reserve prices only, leading to a better welfare performance post response as shown in Figure~\ref{fig:welfare_trend}. Moreover, as the reserve prices become more precise, the average multiplier is lower after convergence. Intuitively, the welfare is maximized when all the value maximizers adopt a bid multiplier $1$ so that the auction results in a ranking according to their values. Therefore, the bid multipliers getting close to $1$ after response is the main driving factor behind the increase of welfare for treatments with reserve prices only. On the other hand, the average multiplier increases during response for both treatments with boosts only, leading to a worse welfare performance post response as shown in Figure~\ref{fig:welfare_trend}. A combination of reserve prices and boosts induces a mild decrease of the average multiplier, resulting in a rather stable welfare performance during value maximizers' response.

\begin{table}[h]
    \centering
    \begin{tabular}{ccccc}
        \toprule
        & Signal & Reserve Only & Boost Only & Boost + Reserve \\
        \midrule
        \multirow{3}{*}{Welfare Lift} 
        & $\gamma = 0.3$ & $37.4\% \pm 3.2\%$ & $23.0\% \pm 3.4\%$ & $55.4\% \pm 3.3\%$ \\
        & $\gamma = 0.5$ & $47.9\% \pm 2.9\%$ & $33.6\% \pm 3.2\%$ & $63.2\% \pm 3.1\%$ \\
        & $\gamma = 0.7$ & $58.6\% \pm 3.3\%$ & $47.7\% \pm 3.2\%$ & $67.6\% \pm 3.2\%$ \\
        \midrule
        \multirow{3}{*}{Revenue Lift} 
        & $\gamma = 0.3$ & $28.2\% \pm 1.7\%$ & $23.7\% \pm 2.2\%$ & $44.9\% \pm 1.9\%$ \\
        & $\gamma = 0.5$ & $39.4\% \pm 1.9\%$ & $35.3\% \pm 2.5\%$ & $52.9\% \pm 1.8\%$ \\
        & $\gamma = 0.7$ & $54.1\% \pm 1.8\%$ & $51.2\% \pm 3.0\%$ & $59.9\% \pm 2.2\%$ \\
        \bottomrule
    \end{tabular}
    \caption{Welfare and revenue lifts for different treatments after convergence.}
    \label{tab:final}
\end{table}

We conduct $10$ runs of repeated experiments and Table~\ref{tab:final} shows the welfare and revenue impact after convergence with $95\%$ confidence intervals. First, notice that treatments with reserve prices only have better welfare and revenue lifts than treatments with boosts when they share the same $\gamma$. Moreover, a combination of boosts and reserve prices leads to further improvements in both revenue and welfare. For all variants, as the signals becomes more precise (i.e., $\gamma$ is closer to $1$), the impacts are larger. Note that in an environment with value maximizers only, if all value maximizers could always hit their target spends after convergence, the improved ratios for welfare and revenue should be the same. However, due to the discontinuity of the bidding landscape, there are value maximizers who cannot hit their targets in practice. As a result, we witness difference between welfare improvement and revenue improvement even after convergence.

\section{Conclusion}

In this paper, we provide both theoretical and empirical evidence to demonstrate that introducing properly chosen reserve prices and/or boosts can improve both revenue and welfare in the auto-bidding world. Our results are robust to bidders' behavior models as well as inaccurate signals that approximate bidders' values. One limitation of our results is regarding the requirement of reserve prices not exceeding bidder values. Although one can avoid this by scaling down the reserve prices, it does require additional knowledge from the auction designer on an upper limit of the inaccurate signals. It is an intriguing question to address such an issue in a more robust way.

\section{Acknowledgement}
Thanks are due to Mohammad Mahdian, Aranyak Mehta, Renato Paes Leme, and Ying Wang for helpful discussions, and for their comments and suggestions.

\bibliographystyle{plainnat}
\bibliography{bib}

\clearpage
\appendix

\section{Missing Proofs of Section \ref{sec:main_lemma}}
\label{sec:main_lemma_app}

\begin{proof}(of Lemma \ref{lem:meta_boost})

For notation convenience, we define $S_{j,k}$ and $S^\opt_{j,k}$ to be the set of bidders who get allocated to any of the top-$k$ slots of auction $j$ in $\A$ and $\opt$, respectively.

The optimal welfare can be written, by exchanging summations, as
\[
\wel(\opt) = \sum_{j=1}^m \sum_{k =1}^{s_j} \sum_{i \in S^{\opt}_{j,k}}(\pos_{j, k} - \pos_{j,k+1}) \cdot v_{i,j},
\]
and the welfare of $\A$ on bids $b$ can be written as
\[
\wel(\A(b)) = \sum_{j=1}^m \sum_{k =1}^{s_j} \sum_{i \in S_{j,k}}(\pos_{j, k} - \pos_{j,k+1}) \cdot v_{i,j}.
\]

We start by lower bounding $\rev(\A(b))$. For each auction $j$, by the condition on payments and rearranging terms, we get 
\[
\rev(\A(b)) \geq \sum_{j=1}^m \sum_{k =1}^{s_j} (\pos_{j, k} - \pos_{j,k+1}) \cdot \sum_{i \in S_{j,k}}\max(\hat{b}_{k+1,j} -z_{i,j}, v_{i,j}\cdot \beta).
\]

For $k$-th slot in auction $j$, we partition the revenue into two components:
\begin{align*}
& \sum_{i \in S_{j,k}}\max(\hat{b}_{k+1,j} -z_{i,j}, v_{i,j}\cdot \beta)\\
 = &  \sum_{i \in S_{j,k} \cap S^\opt_{j,k}}\max(\hat{b}_{k+1,j} -z_{i,j}, v_{i,j}\cdot \beta) +   \sum_{i \in S_{j,k} \backslash S^\opt_{j,k}}\max(\hat{b}_{k+1,j} -z_{i,j}, v_{i,j}\cdot \beta)
\end{align*}
For the first component, we simply have
\[
 \sum_{i \in S_{j,k} \cap S^\opt_{j,k}}\max(\hat{b}_{k+1,j}-z_{i,j} , v_{i,j}\cdot \beta) \geq  \sum_{i \in S_{j,k} \cap S^\opt_{j,k}}v_{i,j}\cdot \beta\numberthis \label{eqn:rev_ine_1}.
\]
For the second component, consider any $i \in  S^\opt_{j,k} \backslash S_{j,k}$. We know in $\A$, bidder $i$ ranks out of top-$k$ in auction $j$. This means $\hat{b}_{k+1,j}$ is at least as high as bidder $i$'s score which is at least $v_{i,j} \cdot \alpha + z_{i,j}$. Since $|S^\opt_{j,k}| = |S_{j,k} | = k$, we know $|S_{j,k} \backslash S^\opt_{j,k}|  = k-|S_{j,k} \cap S^\opt_{j,k}| =| S^\opt_{j,k} \backslash S_{j,k}|$. Therefore,
\begin{align*}
&\sum_{i \in S_{j,k} \backslash S^\opt_{j,k}}\max(\hat{b}_{k+1,j} -z_{i,j}, v_{i,j}\cdot \beta) \\
\geq & |S_{j,k} \backslash S^\opt_{j,k}| \cdot \hat{b}_{k+1,j} - \sum_{i \in S_{j,k} \backslash S^\opt_{j,k}}z_{i,j}\\
=&  | S^\opt_{j,k} \backslash S_{j,k}| \cdot \hat{b}_{k+1,j}- \sum_{i \in S_{j,k} \backslash S^\opt_{j,k}}z_{i,j}\\
\geq& \left(\sum_{i \in  S^\opt_{j,k} \backslash S_{j,k}}v_{i,j} \cdot \alpha + z_{i,j}\right) - \left(\sum_{i \in S_{j,k} \backslash S^\opt_{j,k}}z_{i,j}\right)\\
\geq& \left(\sum_{i \in  S^\opt_{j,k} \backslash S_{j,k}}v_{i,j} \cdot (\alpha +\mu) \right) - \left(\sum_{i \in S_{j,k} \backslash S^\opt_{j,k}}v_{i,j}\cdot \nu\right)\numberthis \label{eqn:rev_ine_2}\\
\end{align*}
We also have 
\[
 \sum_{i \in S_{j,k} \backslash S^\opt_{j,k}}\max(\hat{b}_{k+1,j}-z_{i,j} , v_{i,j}\cdot \beta) \geq   \sum_{i \in S_{j,k} \backslash S^\opt_{j,k}}v_{i,j}\cdot \beta \numberthis \label{eqn:rev_ine_3}.
\]
Therefore, by inequalities \eqref{eqn:rev_ine_2} and \eqref{eqn:rev_ine_3}, we get
\begin{align*}
&\sum_{i \in S_{j,k} \backslash S^\opt_{j,k}}\max(\hat{b}_{k+1,j} -z_{i,j}, v_{i,j}\cdot \beta) \\
= & \frac{\beta}{\beta+\nu}\left(\sum_{i \in S_{j,k} \backslash S^\opt_{j,k}}\max(\hat{b}_{k+1,j} -z_{i,j}, v_{i,j}\cdot \beta)\right) +\frac{\nu}{\beta+\nu}\left(\sum_{i \in S_{j,k} \backslash S^\opt_{j,k}}\max(\hat{b}_{k+1,j} -z_{i,j}, v_{i,j}\cdot \beta)\right)\\
\geq & \left(\sum_{i \in  S^\opt_{j,k} \backslash S_{j,k}}v_{i,j} \cdot (\alpha +\mu) \cdot\frac{\beta}{\beta+\nu}\right) - \left(\sum_{i \in S_{j,k} \backslash S^\opt_{j,k}}v_{i,j}\cdot \nu\cdot\frac{\beta}{\beta+\nu}\right) + \left( \sum_{i \in S_{j,k} \backslash S^\opt_{j,k}}v_{i,j}\cdot \beta\cdot\frac{\nu}{\nu+\beta}\right)\\
&= \sum_{i \in  S^\opt_{j,k} \backslash S_{j,k}}v_{i,j}  \cdot\frac{(\alpha +\mu)\beta}{\beta+\nu}.
\end{align*}
Together with inequality \eqref{eqn:rev_ine_1}, we have
\begin{align*}
    \rev(\A(b)) \geq& \sum_{j=1}^m \sum_{k =1}^{s_j} (\pos_{j, k} - \pos_{j,k+1}) \cdot \sum_{i \in S_{j,k}}\max(\hat{b}_{k+1,j}-z_{i,j} , v_{i,j}\cdot \beta)\\
   \geq& \sum_{j=1}^m \sum_{k =1}^{s_j} (\pos_{j, k} - \pos_{j,k+1})\left(\sum_{i \in S_{j,k} \cap S^\opt_{j,k}}v_{i,j}\cdot \beta+\sum_{i \in  S^\opt_{j,k} \backslash S_{j,k}}v_{i,j} \cdot\frac{(\alpha +\mu)\beta}{\beta+\nu}\right)\\
   \geq& \sum_{j=1}^m \sum_{k =1}^{s_j} (\pos_{j, k} - \pos_{j,k+1})\sum_{i \in S^\opt_{j,k}}v_{i,j} \cdot \min\left(\frac{(\alpha +\mu)\beta}{\beta+\nu}, \beta\right)\\
   =&  \min\left(\frac{(\alpha +\mu)\beta}{\beta+\nu}, \beta\right)\cdot\wel(\opt).
\end{align*}

Now we lower bound $\wel(\A(b))$. By the third condition, we know $\wel(\A(b)) \geq \rev(\A(b))$. Here we prove a slightly different lower bound on $\rev(\A(b))$ using inequalities \eqref{eqn:rev_ine_1} and \eqref{eqn:rev_ine_2}:
\begin{align*}
    &\rev(\A(b))\\ \geq& \sum_{j=1}^m \sum_{k =1}^{s_j} (\pos_{j, k} - \pos_{j,k+1}) \cdot \sum_{i \in S_{j,k}}\max(\hat{b}_{k+1,j}-z_{i,j} , v_{i,j}\cdot \beta)\\
    \geq& \sum_{j=1}^m \sum_{k =1}^{s_j} (\pos_{j, k} - \pos_{j,k+1})\left(\sum_{i \in S_{j,k} \cap S^\opt_{j,k}}v_{i,j}\cdot \beta+\sum_{i \in  S^\opt_{j,k} \backslash S_{j,k}}v_{i,j}\cdot (\alpha+\mu) - \sum_{i \in  S_{j,k} \backslash S^{\opt}_{j,k}}v_{i,j}\cdot \nu\right)\numberthis \label{eqn:rev_ine_4}
\end{align*}
There are two cases. If $\alpha +\mu - \beta \leq \nu$, we have
\begin{align*}
&\wel(\A(b))\\ \geq& \frac{\nu}{1+\nu}\cdot \wel(\A(b)) + \frac{1}{1+\nu} \cdot \rev(\A(b))\\
\geq& \sum_{j=1}^m \sum_{k =1}^{s_j} (\pos_{j, k} - \pos_{j,k+1}) \left(\sum_{i \in S_{j,k} \cap S^\opt_{j,k}}  v_{i,j}+\sum_{i \in  S_{j,k} \backslash S^{\opt}_{j,k}}v_{i,j}\right)\cdot  \frac{\nu}{1+\nu} \\
&+\sum_{j=1}^m \sum_{k =1}^{s_j} (\pos_{j, k} - \pos_{j,k+1})\\
&\cdot\left(\sum_{i \in S_{j,k} \cap S^\opt_{j,k}}v_{i,j}\cdot \beta+\sum_{i \in  S^\opt_{j,k} \backslash S_{j,k}}v_{i,j}\cdot (\alpha+\mu) - \sum_{i \in  S_{j,k} \backslash S^{\opt}_{j,k}}v_{i,j}\cdot \nu\right) \cdot \frac{1}{1+\nu}\\
&\text{(by the definition of $\wel(\A(b))$ and inequality \eqref{eqn:rev_ine_4})} \\
=&\sum_{j=1}^m \sum_{k =1}^{s_j} (\pos_{j, k} - \pos_{j,k+1})\left(\sum_{i \in S_{j,k} \cap S^\opt_{j,k}}v_{i,j}\cdot \frac{\beta+\nu}{1+\nu}+\sum_{i \in  S^\opt_{j,k} \backslash S_{j,k}}v_{i,j}\cdot\frac{\alpha+\mu}{1+\nu} \right) \\
\geq &\frac{\alpha+\mu}{1+\nu}\cdot\sum_{j=1}^m \sum_{k =1}^{s_j} (\pos_{j, k} - \pos_{j,k+1})\left(\sum_{i \in S_{j,k} \cap S^\opt_{j,k}}v_{i,j}+\sum_{i \in  S^\opt_{j,k} \backslash S_{j,k}}v_{i,j} \right) \\
=& \frac{\alpha+\mu}{1 + \max(\nu, \alpha+ \mu -\beta)}\cdot \wel(\opt).
\end{align*}

If $\alpha +\mu - \beta >\nu$, we have
\begin{align*}
&\wel(\A(b))\\ \geq& \frac{\alpha+\mu- \beta}{1+\alpha+\mu- \beta}\cdot  \wel(\A(b))  +\frac{1}{1+\alpha+\mu- \beta}\cdot  \rev(\A)  \\
\geq &\frac{\alpha+\mu}{1+\alpha+\mu - \beta}\cdot\sum_{j=1}^m \sum_{k =1}^{s_j} (\pos_{j, k} - \pos_{j,k+1})\\
&\cdot\left(\sum_{i \in S_{j,k} \cap S^\opt_{j,k}}v_{i,j}+\sum_{i \in  S^\opt_{j,k} \backslash S_{j,k}}v_{i,j}+\sum_{i \in  S_{j,k} \backslash S^{\opt}_{j,k}}v_{i,j} \cdot\frac{\alpha+\mu-\beta-\nu}{\alpha+\mu}\right) \\
&\text{(by the definition of $\wel(\A(b))$ and inequality \eqref{eqn:rev_ine_4})} \\
\geq &\frac{\alpha+\mu}{1+\alpha+\mu - \beta}\cdot\sum_{j=1}^m \sum_{k =1}^{s_j} (\pos_{j, k} - \pos_{j,k+1})\left(\sum_{i \in S_{j,k} \cap S^\opt_{j,k}}v_{i,j}+\sum_{i \in  S^\opt_{j,k} \backslash S_{j,k}}v_{i,j} \right) \\
=& \frac{\alpha+\mu}{1 + \max(\nu, \alpha+ \mu -\beta)}\cdot \wel(\opt).
\end{align*}
\end{proof}
\section{Missing Proofs of Section \ref{sec:vcg}}
\label{sec:vcg_app}

\begin{proof}(of Lemma \ref{lem:vcg})
We prove by contradiction. Suppose that there exists some $b \in \Theta$ such that $b_{i',j'} < v_{i',j'}$ for some $i' \in [n], j' \in [m]$, and $v_{i',j'}$ ranks in top-$s_{j'}$ in auction $j'$. Define $b'_{i'}$ to be the same as $b_{i'}$ except $b'_{i',j'} = v_{i',j'}$. We want to show $b'_{i'}$ dominates $b_{i'}$.

We start with proving the first requirement for $b'_{i'}$ dominating $b_{i'}$. For any $b'_{-i'}$, let $x,p$ be the allocation of bids $(b_{i'}, b'_{-i'})$, and $x',p'$ be the allocation of bids $(b'_{i'}, b'_{-i'})$. Since bids are the same in auctions other than $j'$, we know that $x_{i',j,k} = x'_{i',j,k}$ and $p_{i',j} = p'_{i',j}$ for any $j \neq j'$ and $k\in [s_j]$. Therefore, the payment difference for bidder $i$ is 
\[
\sum_{j =1}^m (p'_{i',j} - p_{i',j}) = p'_{i',j'} -p_{i',j'}.
\] 
and the welfare difference for bidder $i$ is
\[
\sum_{j =1}^m \sum_{k=1}^{s_j}(x'_{i',j,k} - x_{i',j,k})\cdot v_{i',j} \cdot \pos_{j,k}= \sum_{k=1}^{s_{j'}}(x'_{i',j',k} -x_{i',j',k})\cdot v_{i',j'}\cdot \pos_{j',k}
\]
We want to show that
\[
0 \leq p'_{i',j'} -p_{i',j'} \leq \sum_{k=1}^{s_{j'}}(x'_{i',j',k} -x_{i',j',k})\cdot v_{i',j'}\cdot \pos_{j',k}.
\]
There are two cases. The first case is that bidder $i'$ does not get allocated in auction $j'$ in allocation $x$. In this case, we only need to show $0 \leq p'_{i',j'}  \leq \sum_{k=1}^{s_{j'}}x'_{i',j',k}\cdot v_{i',j'}\cdot \pos_{j',k}$. Notice that $b_{i',j'} = v_{i',j'}$. And we can derive from the payment definition of VCG that the payment is non-negative and at most bid multiplied by the position normalizer. Therefore, we have $0 \leq p'_{i',j'}  \leq \sum_{k=1}^{s_{j'}}x'_{i',j',k}\cdot v_{i',j'}\cdot \pos_{j',k}$ .

In the second case, bidder $i'$ gets allocated to some slot $q$ in auction $j'$ in allocation $x$. Since $b'_{i',j'} = v_{i',j'} > b_{i',j'}$, we know bidder $i'$ gets allocated to some slot $q'$ in auction $j'$ in allocation $x$ and we have $q'\leq q$. Now we can write the welfare difference
\[
\sum_{k=1}^{s_{j'}}(x'_{i',j',k} -x_{i',j',k})\cdot v_{i',j'}\cdot \pos_{j',k} = v_{i',j'} \cdot (\pos_{j',q'} - \pos_{j',q}).
\]
On the other hand, the payment difference $p'_{i',j'} -p_{i',j'}$ can be upper bounded by $(\pos_{j',q'} - \pos_{j',q})$ multiplied by the maximum of $r_{i',j'}$ and $v_{i',j'}$. And we also know that $r_{i',j'} \leq v_{i',j'}$. Therefore, 
\[
0 \leq p'_{i',j'} -p_{i',j'} \leq (\pos_{j',q'} - \pos_{j',q}) \cdot v_{i',j'}
\]
To sum up, we get that the payment difference is at most the welfare difference, and they are non-negative, i.e.
\[
0 \leq \sum_{j =1}^m (p'_{i',j} - p_{i',j}) \leq \sum_{j =1}^m \sum_{k=1}^{s_j}(x'_{i',j,k} - x_{i',j,k})\cdot v_{i',j} \cdot \pos_{j,k}.
\]
Since $\lambda_{i'} \in [0,1]$, this implies that the objective is higher in $(x',p')$ than in $(x,p)$, i.e.
\[
 \sum_{j =1}^m \sum_{k=1}^{s_j}x'_{i',j,k} \cdot v_{i',j} \cdot \pos_{j,k} -\lambda_{i'} \cdot \sum_{j =1}^m p'_{i',j} \geq  \sum_{j =1}^m \sum_{k=1}^{s_j}x_{i',j,k} \cdot v_{i',j} \cdot \pos_{j,k} -\lambda_{i'} \cdot \sum_{j =1}^m p_{i',j}.
\]
And the constraint is better satisfied, i.e.
\[
 \sum_{j =1}^m \sum_{k=1}^{s_j}x'_{i',j,k} \cdot v_{i',j} \cdot \pos_{j,k} - \sum_{j =1}^m p'_{i',j} \geq  \sum_{j =1}^m \sum_{k=1}^{s_j}x_{i',j,k} \cdot v_{i',j} \cdot \pos_{j,k} - \sum_{j =1}^m p_{i',j}.
\]
They together imply the first requirement for $b'_{i'}$ dominating $b_{i'}$ is satisfied.

Now we continue to prove the second requirement for $b'_{i'}$ dominating $b_{i'}$. Consider the following $b'_{-i'}$:
\begin{itemize}
    \item For auction $j'$, and bidder $i \neq i'$ with value ranks higher than bidder $i$ in auction $j'$, set $b'_{i,j'} = v_{i,j'} + z_{i',j'}$.
    \item For auction $j'$, and bidder $i \neq i'$ with value ranks lower than bidder $i$ in auction $j'$, set $b'_{i,j'} = \frac{v_{i,j'} + b_{i',j'}}{2} + z_{i',j'} - z_{i,j'}$.
    \item For other auction $j \neq j'$ and bidder $i\neq i'$, set $b'_{i,j} = 0$.
\end{itemize}
Again, let $x,p$ be the allocation of bids $(b_{i'}, b'_{-i'})$, and $x',p'$ be the allocation of bids $(b'_{i'}, b'_{-i'})$. For auction $j'$, let bidder $i'$'s value ranks at $q$. With bid $b'_{i',j'} = v_{i',j'}$, it's easy to check that bidder $i'$'s score ranks also at $q$. On the other hand, with bid $b_{i',j'} < v_{i',j'}$, bidder $i'$ ranks the last. So in $x'$ the allocation is better and it's easy to see that the welfare improvement in $x'$ is larger than the payment increase. For other auction $j \neq j'$, bidder $i'$ gets the same allocation and payments in $(x,p)$ and $(x',p')$. Therefore, we know
\[
\wel_{i'}(x',p') -\lambda_{i'} \cdot \rev_{i'}(x',p') > \wel_{i'}(x,p)-\lambda_{i'} \cdot \rev_{i'}(x,p).
\]
In auction $j \neq j'$, competing bids are 0 and therefore, we bound payments by reserves: 
\[
p'_{i',j}\leq \sum_{k=1}^{s_j} x'_{i',j,k} \cdot r_{i',j} \cdot \pos_{j,k}\leq \sum_{k=1}^{s_j} x'_{i',j,k} \cdot v_{i',j}.
\]
In auction $j'$, we bound payments by the bid $b'_{i',j'} = v'_{i',j'}$: \[
p'_{i',j'}\leq \sum_{k=1}^{s_j'} x'_{i',j',k} \cdot b_{i',j'} \cdot \pos_{j',k}=  \sum_{k=1}^{s_j'} x'_{i',j',k} \cdot v_{i',j'}.
\]
They together give $\rev_{i'}(x',p')\leq \wel_{i'}(x',p')$. 

With $\wel_{i'}(x',p') -\lambda_{i'} \cdot \rev_{i'}(x',p') > \wel_{i'}(x,p)-\lambda_{i'} \cdot \rev_{i'}(x,p)$ and $\rev_{i'}(x',p')\leq \wel_{i'}(x',p')$, we know the second requirement for $b'_{i'}$ dominating $b_{i'}$ is satisfied.
Now we have $b'_{i'}$ dominates $b_{i'}$. Thus, we get a contradiction.
\end{proof}
\section{Missing Proofs of Section \ref{sec:gsp}}
\label{sec:gsp_app}

\begin{proof}(of Lemma \ref{lem:gsp_u})
We prove by contradiction. Suppose that there exists some uniform bids $b \in \Theta_u$ such that $b_{i',j'} <v_{i',j'}$ for some $i' \in [n], j' \in [m]$. Since $b_{i'}$ is a uniform bidding, we can write $b_{i',j} = \delta_{i'} \cdot v_{i',j}, \forall j\in[m]$,  and we know $\delta_{i'} < 1$.

Define $b'_{i'}$ to be a uniform bidding such that $b'_{i',j} = v_{i',j},\forall j \in [m]$. We want to show $b'_{i'}$ dominates $b_{i'}$.

We start with proving the first requirement for $b'_{i'}$ dominating $b_{i'}$. For any $b'_{-i'}$, let $x,p$ be the allocation of bids $(b_{i'}, b'_{-i'})$, and $x',p'$ be the allocation of bids $(b'_{i'}, b'_{-i'})$. Since $b_{i',j} \geq b'_{i',j}, \forall j \in[m]$, we know that bidder $i'$ is ranked no worse in $x$ than in $x'$ for each auction $j \in [m]$. This implies that 
\[
\wel_{i'}(x',p') \geq \wel_{i'}(x,p) 
\]
On the other hand, we know that in GSP with reserve, the payment in each auction for each bidder is at most its bid multiplied by the position normalizer. Therefore, we have 
\[
\rev_{i'}(x',p') \leq \sum_{j =1}^m \sum_{k=1}^{s_j}x'_{i',j,k}\cdot b'_{i',j} \cdot \pos_{j,k}  \leq  \sum_{j =1}^m \sum_{k=1}^{s_j}x'_{i',j,k}\cdot v_{i',j} \cdot \pos_{j,k} =\wel_{i'}(x',p').
\]

Now we continue to prove the second requirement for $b'_{i'}$ dominating $b_{i'}$. Consider the following $b'_{-i'}$:
\begin{itemize}
    \item For auction $j'$, and bidder $i \neq i'$ with value ranks higher than bidder $i$ in auction $j'$, set $b'_{i,j'} = v_{i,j'} + z_{i',j'}$.
    \item For auction $j'$, and bidder $i \neq i'$ with value ranks lower than bidder $i$ in auction $j'$, set $b'_{i,j'} = \frac{v_{i,j'} + b_{i',j'}}{2} + z_{i',j'} - z_{i,j'}$.
    \item For other auction $j \neq j'$ and bidder $i\neq i'$, set $b'_{i,j}$ to be consistent with $b'_{i,j'}$ according to uniform bidding.
\end{itemize}
Again, let $x,p$ be the allocation of bids $(b_{i'}, b'_{-i'})$, and $x',p'$ be the allocation of bids $(b'_{i'}, b'_{-i'})$. For auction $j'$, let bidder $i'$'s value ranks at $q$. With bid $b'_{i',j'} = v_{i',j'}$, it's easy to check that bidder $i'$'s score ranks also at $q$. On the other hand, with bid $b_{i',j'} < v_{i',j'}$, bidder $i'$ ranks the last. So in $x'$ the allocation is better and it's easy to see that the welfare improvement in $x'$ is larger than the payment increase. For other auction $j \neq j'$, bidder $i'$ gets the same allocation and payments in $(x,p)$ and $(x',p')$. Therefore, we know
\[
\wel_{i'}(x',p') -\lambda_{i'} \cdot \rev_{i'}(x',p') > \wel_{i'}(x,p)-\lambda_{i'} \cdot \rev_{i'}(x,p).
\]
And we have already shown $\rev_{i'}(x',p')\leq \wel_{i'}(x',p')$ for the first requirement. 

With $\wel_{i'}(x',p') -\lambda_{i'} \cdot \rev_{i'}(x',p') > \wel_{i'}(x,p)-\lambda_{i'} \cdot \rev_{i'}(x,p)$ and $\rev_{i'}(x',p')\leq \wel_{i'}(x',p')$, we know the second requirement for $b'_{i'}$ dominating $b_{i'}$ is satisfied.
Thus, we have $b'_{i'}$ dominates $b_{i'}$. And we get a contradiction.
\end{proof}

\begin{proof}(of Lemma \ref{lem:gsp})
We prove by contradiction. Suppose that there exists some $b \in \Theta$ such that $b_{i',j'} <r_{i',j'}$ for some $i' \in [n], j' \in [m]$. Define $b'_{i'}$ to be the same as $b_{i'}$ except $b'_{i',j'} = r_{i',j'}$. We want to show $b'_{i'}$ dominates $b_{i'}$.

We start with proving the first requirement for $b'_{i'}$ dominating $b_{i'}$. For any $b'_{-i'}$, let $x,p$ be the allocation of bids $(b_{i'}, b'_{-i'})$, and $x',p'$ be the allocation of bids $(b'_{i'}, b'_{-i'})$. Since bids are the same in auctions other than $j'$, we know that $x_{i',j,k} = x'_{i',j,k}$ and $p_{i',j} = p'_{i',j}$ for any $j \neq j'$ and $k\in [s_j]$. Since $b_{i',j'} <  r_{i',j'}$, we know bidder $i'$ is not allocated in auction $j'$ in $x$. So $p_{i',j'} =0$ and $x_{i',j',k} =0, \forall k \in [s_{j'}]$. Therefore, the payment difference for bidder $i$ is 
\[
\sum_{j =1}^m (p'_{i',j} - p_{i',j}) = p'_{i',j'} .
\] 
and the welfare difference for bidder $i$ is
\[
\sum_{j =1}^m \sum_{k=1}^{s_j}(x'_{i',j,k} - x_{i',j,k})\cdot v_{i',j} \cdot \pos_{j,k}= \sum_{k=1}^{s_{j'}}x'_{i',j',k}\cdot v_{i',j'}\cdot \pos_{j',k}.
\]

We want to show
\[
0 \leq p'_{i',j'}  \leq \sum_{k=1}^{s_{j'}}x'_{i',j',k}\cdot v_{i',j'}\cdot \pos_{j',k}.
\]
There are two cases. The first case is that bidder $i'$ does not get allocated in auction $j'$ in allocation $x'$. In this case, we simply have 
\[
0 =p'_{i',j'}  = \sum_{k=1}^{s_{j'}}x'_{i',j',k}\cdot v_{i',j'}\cdot \pos_{j',k}.
\]

In the second case, bidder $i'$ gets allocated to some slot $q'$ in auction $j'$ in allocation $x'$. We can write $\sum_{k=1}^{s_{j'}}x'_{i',j',k}\cdot v_{i',j'}\cdot \pos_{j',k}$ as $\pos_{j',q'} \cdot v_{i',j'} \cdot \gamma$. Since in GSP with reserve, the payment in each auction for each bidder is at most the bid multiplied by the position normalizer and we have bid $b'_{i',j'} =  r_{i',j'}$, we get
\[
0 \leq p'_{i',j'}  \leq  \sum_{k=1}^{s_{j'}}x'_{i',j',k}\cdot r_{i',j'}\cdot \pos_{j',k}\leq  \sum_{k=1}^{s_{j'}}x'_{i',j',k}\cdot v_{i',j'}\cdot \pos_{j',k}.
\]
Now we have 

\[
0 \leq \sum_{j =1}^m (p'_{i',j} - p_{i',j}) \leq \sum_{j =1}^m \sum_{k=1}^{s_j}(x'_{i',j,k} - x_{i',j,k})\cdot v_{i',j} \cdot \pos_{j,k}.
\]
The rest of the proof follows exactly as the proof for the first requirement in Lemma \ref{lem:vcg}. 

Now we proceed to prove the second requirement for $b'_{i'}$ dominating $b_{i'}$. Consider the simple $b'_{-i'}$ with all zeroes. Again, let $x,p$ be the allocation of bids $(b_{i'}, b'_{-i'})$, and $x',p'$ be the allocation of bids $(b'_{i'}, b'_{-i'})$. For auction $j'$, clearly bidder $i'$ gets allocated a slot in $x'$ but not in $x$ due to $b_{i',j'} < r_{i',j'}$. And in $(x',p')$, this slot price is the reserve multiplied by the position normalizer which is lower than the value multiplied by the position normalizer. For other auction $j \neq j'$, bidder $i'$ gets the same allocation and price in $(x,p)$ and $(x',p')$. Therefore, we know
\[
\wel_{i'}(x',p') -\lambda_{i'} \cdot \rev_{i'}(x',p') > \wel_{i'}(x,p)-\lambda_{i'} \cdot \rev_{i'}(x,p) .
\]
Since bidders other than bidder $i'$ bid all zeroes and there are no boosts, bidder $i'$ payment can be bounded by reserves, i.e.
\[
\rev_{i'}(x',p') \leq \sum_{j=1}^m p'_{i',j} = \sum_{j =1}^m \sum_{k=1}^{s_j} x'_{i',j,k} \cdot r_{i',j} \cdot \pos_{j,k}\leq \sum_{j =1}^m \sum_{k=1}^{s_j} x'_{i',j,k} \cdot v_{i',j} \cdot \pos_{j,k} = \wel_{i'}(x',p').
\]
With $\wel_{i'}(x',p') -\lambda_{i'} \cdot \rev_{i'}(x',p') > \wel_{i'}(x,p)-\lambda_{i'} \cdot \rev_{i'}(x,p)$ and $\rev_{i'}(x',p')\leq \wel_{i'}(x',p')$, we know the second requirement for $b'_{i'}$ dominating $b_{i'}$ is satisfied.

We get $b'_{i'}$ dominates $b_{i'}$. And then we get a contradiction.
\end{proof}
\section{Missing Proofs of Section~\ref{sec:fpa}}
\label{app:fpa}

\begin{proof}(of Lemma \ref{lem:fpa})
Notice that in the proof Lemma \ref{lem:gsp}, the only property we use about GSP with reserve is that the payment in each auction for each bidder is at most its bid multiplied by the position normalizer. This property also holds in FPA with reserve. Therefore, Lemma \ref{lem:fpa} simply follows from the proof of Lemma \ref{lem:gsp}.
\end{proof}

\section{Instances with Matching Approximation Lower Bounds}
\label{sec:tight_app}

First of all, for revenue approximation ratio, it is tight even for a setting with a single bidder and a single auction. Assume the bidder’s value is $1$ and the signal is between $\gamma$ and $1$. The seller cannot set a reserve larger than the signal; or otherwise when the signal is $1$, the reserve would be larger than the bidder’s value, leading to a 0-approximation. Therefore, when the signal is $\gamma$, the reserve would be at most $\gamma$, leading to a $\gamma$-approximation in revenue.

For welfare approximation ratios, we show three two-bidder examples for the settings with reserve only, boost only, and reserve \& boost. In all examples, buyers’ targets are $1$ and there are two auctions. The values of bidder $2$ are always $0$ for auction $1$ and $1$ for auction $2$. We use different valuations of bidder $1$ in both auctions to establish the lower bounds. $\epsilon > 0$ represents an arbitrarily small number.

\subsection{Tight instance for reserve only auctions}
\begin{center}
\begin{tabular}{c|c|c}
             & Values in Auction $1$ & Values in Auction $2$  \\
  \hline
  Bidder $1$ & $1/(1-\gamma)$        & $\epsilon$  \\
  \hline
  Bidder $2$ & $0$                   & $1$  \\
\end{tabular}
\end{center}

In the above example, the reserve signal for bidder $1$ in auction $1$ is $s\in[\gamma/(1-\gamma),1/(1-\gamma)]$. When the realized signal $s=\gamma/(1-\gamma)$, the reserve is at most $s$, and bidder $1$ has incentive to win both auctions, paying at most $\gamma/(1-\gamma)+1=1/(1-\gamma)$, receiving $1/(1-\gamma)+\epsilon$ total value. This leads to $(1/(1-\gamma)+\epsilon)/(1/(1-\gamma)+1)=1/(2-\gamma)$ in welfare approximation. 

Moreover, as the signal could be as high as bidder $1$’s value in auction $1$, the reserve multiplier cannot be larger than $1$; or otherwise, bidder $1$ would choose to skip the first auction. By using a reserve multiplier that is at most $1$, there is a realization of the signal in which the reserve is at most $\gamma/(1-\gamma)$.

\subsection{Tight instance for boost only auctions}
\begin{center}
\begin{tabular}{c|c|c}
             & Values in Auction $1$ & Values in Auction $2$  \\
  \hline
  Bidder $1$ & $1-\gamma+\epsilon$   & $\gamma$  \\
  \hline
  Bidder $2$ & $0$                   & $1$  \\
\end{tabular}
\end{center}

When there is no boost, bidder $1$ has the incentive to win both auctions, paying $1$ and receiving $(1-\gamma+\epsilon)+\gamma=1+\epsilon$ total value. This leads to an approximation ratio $1/(2-\gamma)$ in welfare.

In the worst case scenario, consider that the boost signals in auction $2$ are $s=\gamma$ for both bidders; and thus, boosts are not effective in changing the auction outcome for auction $2$.

\subsection{Tight instance for auctions with both reserves \& boost}
\begin{center}
\begin{tabular}{c|c|c}
             & Values in Auction $1$ & Values in Auction $2$  \\
  \hline
  Bidder $1$ & $1+\epsilon$          & $\gamma$  \\
  \hline
  Bidder $2$ & $0$                   & $1$  \\
\end{tabular}
\end{center}

When there is no boost and the reserve for bidder $1$ in auction $1$ is at most $\gamma$ (realized signal $s=\gamma$), bidder $1$ has the incentive to win both auctions, paying at most $\gamma+1$ and receiving $(1+\epsilon)+\gamma$ total value. This leads to an approximation ratio $(1+\gamma)/2$ in welfare. In the worst case scenario, consider that the boost signals in auction $2$ are $s=\gamma$ for both bidders; and thus, boosts are not effective in changing the auction outcome for auction $2$. As for the reserves, similar to our analysis in the first example, the seller needs to set a reserve that is at most $\gamma$ for bidder $1$ in auction $1$ for some instances.

\end{document}